\documentclass[twocolumn]{aastex63}
\usepackage{natbib}
\usepackage{color}
\usepackage[colorinlistoftodos]{todonotes}

\gdef\co{CO(3-2) }
\gdef\msun{$M_\odot$}
\gdef\mstar{$M_\star$}

\gdef\kms{km/s}

\graphicspath{{./}{}}

 \received{Sep 27}
 \revised{Oct 19}
 \accepted{Accepted in ApJL Oct 21, 2019}

\shorttitle{Anomalously narrow CO linewidths of cSFGs at $z\sim2.3$}
\shortauthors{Mowla et al.}

\begin{document}

\title{Anomalously Narrow Linewidths of Compact Massive Star-Forming Galaxies at $z\sim2.3$: A Possible Inclination Bias in the Size--Mass Plane}


\author{Lamiya A. Mowla}
\email{lamiya.mowla@yale.edu}
\affil{Astronomy Department, Yale University, New Haven, CT 06511, USA}

\author{Erica J. Nelson}
\thanks{Hubble Fellow}
\affil{Harvard-Smithsonian Center for Astrophysics, 60 Garden Street, Cambridge, MA}

\author{Pieter van Dokkum}
\affil{Astronomy Department, Yale University, New Haven, CT 06511, USA}

\author{Ken-ichi Tadaki}
\affil{National Astronomical Observatory of Japan, 2-21-1 Osawa, Mitaka, Tokyo 181-8588, Japan}

\begin{abstract}
Compact, massive star forming galaxies at $z\sim2.5$ are thought to be building the central regions of giant elliptical galaxies today. However, a significant fraction of these objects were previously shown to have much smaller H$\alpha$ line widths than expected. A possible interpretation is that H$\alpha$ emission from their central regions, where the highest velocities are expected, is typically obscured by dust. Here we present ALMA observations of the CO(3-2) emission line of three compact, massive galaxies with H$\alpha$ line widths of FWHM(H$\alpha$)$\sim$125-260\,km\,s$^{-1}$ to test this hypothesis. 
Surprisingly, in all three galaxies, the CO line width is similar to the H$\alpha$ line width: we find FWHM(CO)$\sim$165\,km\,s$^{-1}$ for all three galaxies whereas FWHM(CO)$\sim$450-700\,km\,s$^{-1}$ was expected from a simple virial estimator. These results show that the narrow H$\alpha$ linewidths of many compact massive star-forming galaxies are not due to preferential obscuration of the highest velocity gas. An alternative explanation for the narrow line widths is that the galaxies are disks that are viewed nearly face-on. We suggest  that there may be an inclination bias in the size-mass plane, such that the apparent rest-frame optical sizes of face-on galaxies are smaller than those of edge-on galaxies. Although not conclusive, this hypothesis is supported by an observed anti-correlation between size and axis ratio of massive galaxies.
\end{abstract}


\section{Introduction} \label{sec:intro}

\begin{figure*}[ht]
    \centering
    \includegraphics[width = \textwidth]{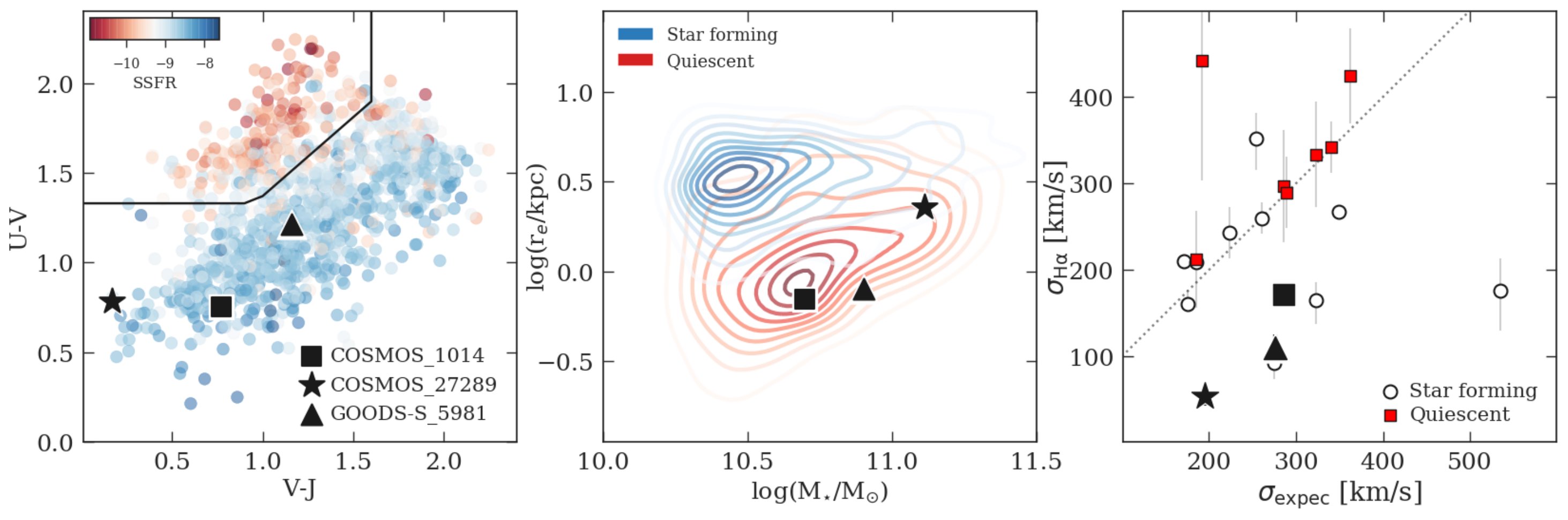}
    \caption{Selection of the sample. \textit{Left: }Distribution of galaxies with $\log$(\mstar/\msun) $>10.3$ and $2.0<z<2.5$ in the $UVJ$ plane, color-coded by the logarithm of the specific star formation rate. \textit{Center: }Size-mass distribution of same sample of galaxies, separated using the $UVJ$ diagram. \textit{Right: }Comparison of observed velocity dispersion (from H$\alpha$ linewidths) and predicted velocity dispersions (calculated using Eq. \ref{eq:sigpred}) of galaxies. Circles show the 14 star-forming compact massive galaxies with no X-ray counterpart from \citet{VanDokkum2015}. Red squares represent velocity dispersions of compact quiescent galaxies at $z\sim2-2.5$ from \citet{VandeSande2011} and \citet{Belli2014}. The three galaxies observed in this study are shown by the black star, square and triangle in all three panels.}
    \label{fig1}
\end{figure*}

The first ``red and dead'' quiescent galaxies appear in redshift surveys as early as $z\sim 3-4$ \citep[e.g.,][]{Glazebrook2017}, and they became the dominant population among massive galaxies by $z\sim1.5-2$ \citep{Brammer2011,Muzzin2013}.
Quiescent galaxies at $z\sim 2$ are remarkably compact, with sizes a factor of $3-4$ smaller
than galaxies of the same mass today \citep[e.g.,][]{Daddi2005,VanDokkum2010,VanderWel2014}.
Their densities match those of the centers of massive elliptical galaxies today, with velocity dispersions as high as $\sigma \sim 300$\,km\,s$^{-1}$
(FWHM\,$\sim 700$\,km\,s$^{-1}$) \citep{Belli2014,VandeSande2011}.
They  are thought to evolve into ellipticals by gradually acquiring outer envelopes through minor mergers \citep{Bezanson2009,Naab2009,VanDokkum2010}.


The immediate progenitors of these compact quiescent galaxies are thought to be
compact highly star-forming galaxies, which have been identified at similar redshifts of $z\sim2-2.5$ 
\citep{Barro2014,Nelson2014,VanDokkum2015}. Although most star forming galaxies at these redshifts are larger than the quiescent population, there is a significant tail of objects with stellar masses $\sim 10^{11}$ \mstar, star formation rates $\sim$ 100 \mstar/yr, and small half-light radii of $\sim$1 kpc \citep{Wuyts2011,Barro2014,Barro2017a}. Their ionized gas, as traced by the H$\alpha$ line, has been found to be in rotating disks that appear to be more extended than the detected stellar light \citep[e.g.,][]{VanDokkum2015,Wisnioski2018}. 

In this context two aspects of these compact massive star forming galaxies are puzzling. The first is that their axis ratio distribution is not consistent with thin disks viewed from random orientations: there are virtually no galaxies in this class with axis ratios $b/a<0.5$ \citep{VanDokkum2015}. The second is that a significant fraction of these galaxies have surprisingly narrow H$\alpha$ line widths.
In a compilation of H$\alpha$ line widths of 14 compact massive star-forming galaxies with no X-ray detection in the CANDELS/3D-HST fields only one galaxy has FWHM\,$>700$\,km\,s$^{-1}$ \citep{Nelson2014}, whereas nine have 120\,$<$\,FWHM\,$<$\,350\,km\,s$^{-1}$. \citep[][hereafter vD15]{VanDokkum2015}.

vD15 attributed the large axis ratios to intrinsically-thick disks and the small
observed line widths to dust obscuration, in combination with falling rotation curves. As the central regions are most obscured the highest velocities may be heavily suppressed in the light of H$\alpha$ \citep{Nelson2016}.
This interpretation is supported by the high dust content of massive $z\sim 2$ galaxies \citep{Whitaker2017}, the measured spatial extent of the H$\alpha$ gas distributions (vD15,\citet{Wisnioski2018}), and independent dynamical evidence for thick disks and falling rotation curves in this mass and redshift range \citep[e.g.,][]{Genzel2017,Lange2016,Ubler2017}.


In this {\em Letter} we test this hypothesis.
Sub-mm observations do not suffer from dust obscuration, and can provide information on the (molecular) gas kinematics in the central regions \citep[e.g.,][]{Tacconi2008,Tadaki2017b,Popping2017,Barro2017b}.  
We used the Atacama Large Millimeter/submillimeter Array (ALMA) to observe three galaxies at $z \sim 2.3$ from the vD15 sample that have very small H$\alpha$ line widths, to determine whether their CO $J=3-2$ line widths are significantly larger.
We assume a Chabrier initial mass function \citep[IMF][]{Chabrier2003} and adopt cosmological parameters of H$_0$ $=$70 \,km\,s$^{-1}$/Mpc, $\Omega_M=$0.3, and $\Omega_{\Lambda}=$0.7.


\section{Sample Selection}




We describe three galaxies objectively selected from a sample of compact star-forming galaxies with H$\alpha$ measurements presented in vD15.
The parent sample was selected from catalogs of the 3D-HST survey \citep{Skelton2014} which utilized photometric and grism spectroscopic information to measure redshifts, rest-frame colors and emission line strengths using the EAZY code \citep{Brammer2011}. Star formation rates were derived from Spitzer MIPS 24$\mu$m data \citep{Whitaker2012, Whitaker2014}, structural parameters were measured from CANDELS imaging \citep{VanderWel2012,VanderWel2014} and stellar masses were determined from fits of stellar population synthesis models to 0.3-8 $\mu$m photometry using the FAST code \citep{Kriek2009}, with Chabrier IMF \citep{Chabrier2003}, Calzetti dust attenuation law \citep{Calzetti2000} and exponentially declining star formation histories. 
The compact star-forming galaxies were selected by vD15 to have
\begin{equation}
    \log (r_e/\rm kpc) < \log (M_{\star}/M_{\odot}) - 10.7,
\end{equation}
where $r_e$ is the circularized effective radius and $M_{\star}$ is the stellar mass. K-band spectroscopy of 25 of these galaxies was obtained using the  MOSFIRE and NIRSPEC spectrographs on Keck. Out of the 25, 14 galaxies have no X-ray counterparts and likely do not host an AGN. The measured velocity dispersions of the galaxies (H$\alpha$ linewidths) compared to their expected velocity dispersions are shown in the right panel of Figure 1. The expected velocity dispersions are calculated using 
\begin{equation}
    \label{eq:sigpred}
    \log (\sigma_{\rm expec}) = 0.5 (\log \rm G + \log \beta(n) + \log(M_{\star}) - \log (r_e)),
\end{equation}
where $\beta$(n) $=$ 8.87 - 0.831 $n$ + 0.0241 $n^2$. Here $n$ is the S\`ersic index and G $=$ 4.31 $\times$ 10$^{-6}$ when $M_{\star}$ is in \msun, $\sigma_{\rm expec}$ is in\,km\,s$^{-1}$ and r$_e$ is in kpc (see vD15). Among the 14 non-AGN galaxies, 9 galaxies have observed H$\alpha$ linewidths lower than $\sigma_{\rm expec}$. For this $Letter$, we selected galaxies with \begin{equation}
    \log \sigma_{\rm exp} - \log \sigma_{\rm obs} > 0.2
\end{equation}
and with axis-ratio $b/a < 0.85$, to ensure that inclination corrections are less than a factor of two. Five out of the nine objects match this criterion. Three of these objects were observable from Chile and formed our sample. These galaxies have stellar masses $\sim$10$^{11}$ \mstar, star formation rates $\sim$200 \mstar/year and circularized half-light radii $\sim$1 kpc. The measured H$\alpha$ FWHMs of the galaxies are 1.5--3.5 times smaller than the expected FWHMs. Properties of the galaxies are shown in Figure \ref{fig1} and listed in Table \ref{tab:1}.
\begin{figure*}[ht]
    \centering
    \includegraphics[width = \textwidth]{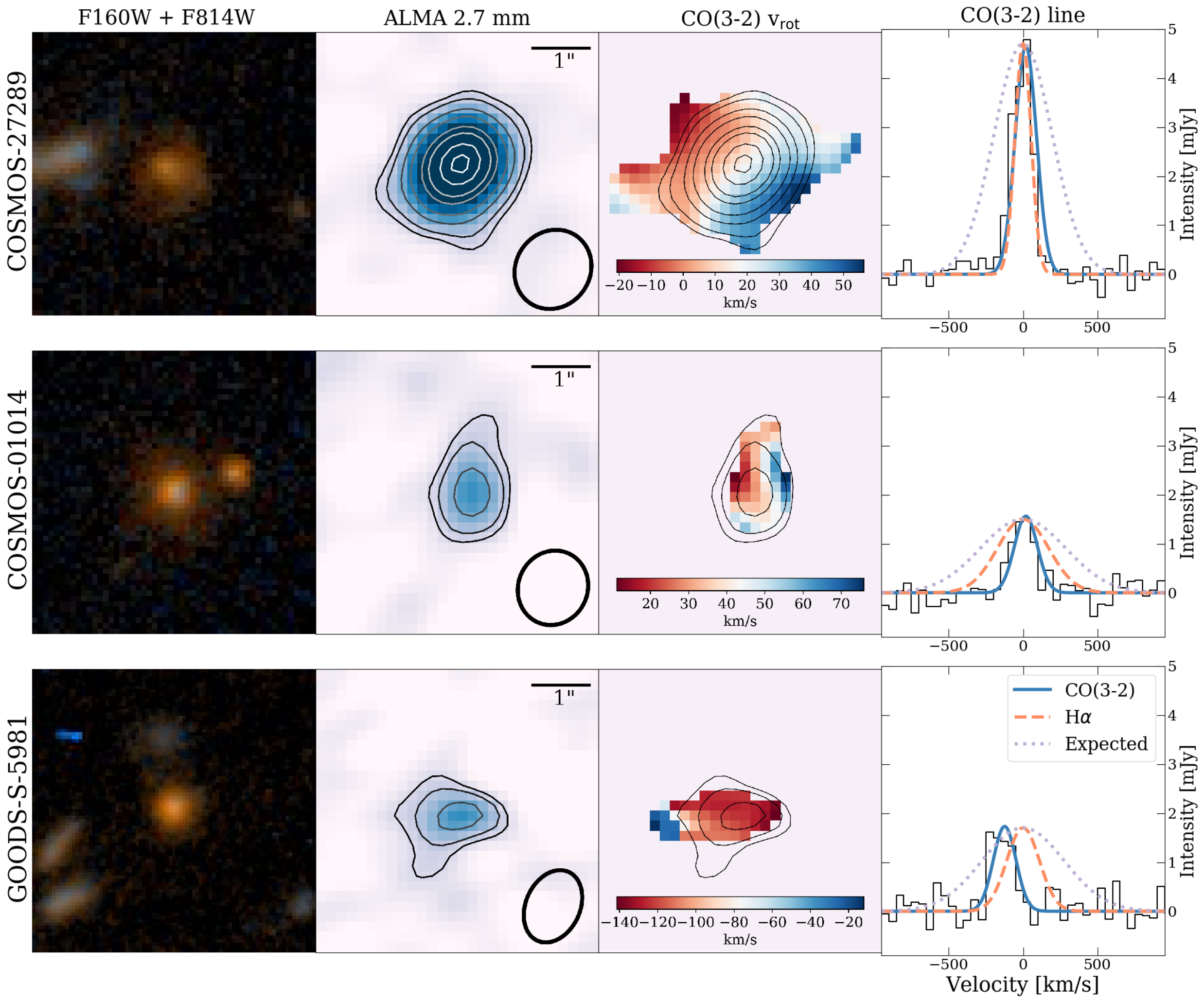}
    \caption{{First column: }Images (5" $\times$ 5"; 40 kpc $\times$ 40 kpc) of COSMOS-27289, COSMOS-01014 and GOODS-S-5981 in WFC3/F160W and ACS/F814W. \textit{Second column: }ALMA CO(3-2) moment-0 maps. \textit{Third column: }CO(3-2) moment-1 maps. \textit{Fourth column: }Spectral profiles of the CO(3-2) line extracted from the ALMA images using 1\farcs5 diameter circular aperture (black line). Best fit single Gaussian fits are shown by the blue line. The measured H$\alpha$ linewidths are shown in orange dashed-lines, and the expected linewidths, calculated from Eq. \ref{eq:sigpred}, are shown by dotted grey lines, both of which are normalized to the peak of the CO flux. }
    \label{fig:kinematics}
\end{figure*}

\begin{table*}[tbh]
	\begin{center}
		\caption{Properties of the three galaxies.}
		\begin{tabular}{cccccccccc}
			\hline \hline \noalign {\smallskip}
			ID & $z$ & SFR & $\log$(\mstar) & $R_e$ & $b/a$ & LW(H$\alpha$)\footnote{All linewidths given as full width half maximums.} &
			LW(pred) & \textbf{LW(\co)} & L'$_{CO(3-2)}$\\
			&     & [\msun/yr] &  & kpc & &\,km\,s$^{-1}$ &\,km\,s$^{-1}$ &\,km\,s$^{-1}$ & 10$^{10}$ K\,km\,s$^{-1}$ pc$^2$ \\
			\hline \hline \noalign {\smallskip}
			COSMOS-27289  & 2.23 & 400 & 11.0 & 2.3 & 0.81 & 127$\pm$28 & 460 & {\bf 167$\pm$7} & 2.39 \\
			COSMOS-1014 & 2.10 &  150 & 10.7  & 0.7 & 0.79 & 400$\pm$30 &674 & {\bf 181$\pm$22} & 0.56 \\
			GOODSS-5981   & 2.25 & 210 & 10.8 & 0.8 & 0.85 & 260$\pm$38 &650 & \textbf{177$\pm$26} & 0.84 \\
			\hline \hline
		\end{tabular}
	\end{center}
	\label{tab:1}
\end{table*}

\section{ALMA observations and Results}
\subsection{Observations and processing}


The three selected galaxies were observed with the \textit{Atacama Large Millimeter/submillimeter Array} (ALMA) as part of ID: 2018.1.01841.S program. The observations were carried out in Band-3 using four spectral windows covering the rest-frame frequency range of $304-349$ GHz. The on-source integration times were $30-42$ minutes in array configuration C43-3 (shortest and longest baselines were 15 and 1240 m, respectively). The water vapor during the observations $PWV=1.67-2.25$ mm. Flux, phase, bandpass and WVR calibrators were also obtained for a total time of $\sim 1$ hr per object. The data were processed through the Common Astronomy Software Application package \citep[CASA;][]{Mcmullin2007}. We use the \texttt{tclean} task with natural weighting to make channel maps and dirty continuum maps excluding the frequency range of the CO line. The spectral resolution of the data is $1-2$\,km\,s$^{-1}$; to search for lines and the continuum the data were binned to a velocity width of 50\,km\,s$^{-1}$. The continuum is undetected or negligible for all three objects. The synthesized beam sizes are $\sim 1\farcs2\times1\farcs2$ and the rms levels are 0.20--0.25 mJy/beam over 50\,km\,s$^{-1}$ in the channel maps. 

\subsection{CO(3-2) Line Detection}

We robustly detect the CO $J=3-2$ emission line in all three galaxies in the expected frequency range from their redshift. We measure the total flux densities of the three galaxies from moment zero maps created using CASA task \texttt{immoment} within the velocity range of -200 to 200\,km\,s$^{-1}$ centered on the CO line. The velocity range is chosen based of preliminary inspection of the data cube; we verified that our results do not change if we increase the velocity range. For C1 and G5 we use apertures of diameter 2", while for C2 we use 3" aperture since it is spatially more extended, as determined from the growth curve. We estimate the line luminosity from the \co line luminosity using
\begin{equation}
    L'_{CO} [\textrm{K\,km\,s$^{-1}$ \rm pc$^2$}] = 3.25 \times 10^7 (S_{CO} \Delta v) \frac{D^2_L}{(1+z)^3 \nu^2_{\rm obs}}
\end{equation}
where $S_{CO} \Delta v$ is the line flux and $D_L$ is the luminosity distance. The resulting values are listed in Table 1. It should be be noted that for G5 the center of CO(3-2) line is -120\,km\,s$^{-1}$ offset from the H${\alpha}$ line. This may suggest that the H${\alpha}$ and CO(3-2) are tracing different components of the gas.


\subsection{CO(3-2) linewidth}
Due to their large masses and small sizes compact star forming galaxies should have large line widths. 
We measure the integrated line widths of the galaxies by fitting a single Gaussian to the spectra, extracted with a 1\farcs5 diameter circular aperture (the results are not sensitive to the size of the aperture). 

The expected linewidths for these galaxies based on their masses, sizes, and geometries are 450--670\,km\,s$^{-1}$ (see Eq. \ref{eq:sigpred}).
Instead we observe 167\,km\,s$^{-1}$, 177\,km\,s$^{-1}$ and 179\,km\,s$^{-1}$ (for C2, C1, and G5 respectively). 
These values are three times smaller than the expected CO line widths and the linewidth of CO is even narrower than that of H$\alpha$ in two of the three galaxies.


The fact that velocity dispersion of molecular gas is similar to (or even lower than) that of H$\alpha$ emission rules out dust obscuration at high velocity gas being responsible for the low H$\alpha$ linewidths. Instead it may indicate that these galaxies are nearly face-on disks; we explore this further by investigating their spatially resolved kinematics.  

\subsection{Spatially-resolved kinematics}

\begin{figure*}[ht]
    \centering
    \includegraphics[width = 0.8\textwidth]{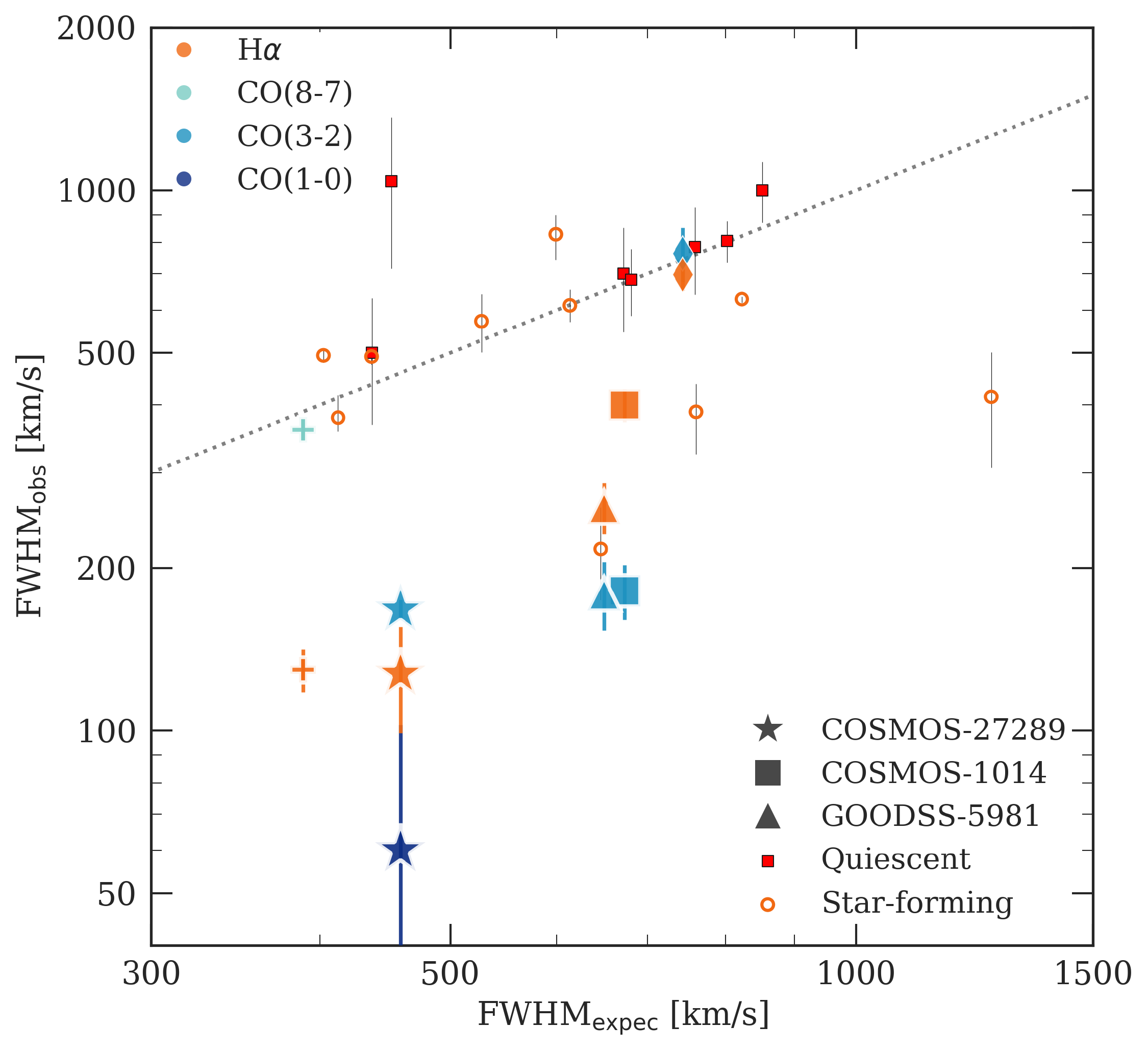}
    \caption{Measured and expected FWHMs of galaxies. Different colors represent observations of different molecular gas tracers. Circles are H$\alpha$ linewidths of cSFGs from vD15 without X-ray counterparts. The red squares represent velocity dispersions of compact quiescent galaxies at $z\sim2-2.5$ from \citet{VandeSande2011,Belli2014}. The star, triangle and square represent COSMOS-27289, COSMOS-01014 and GOODS-S-5981 respectively, the three galaxies studied here from vD15. The CO (1-0) linewidth of COSMOS-27289 is from \citep{Spilker2016} (dark blue star). We have also added the FWHMs of GOODS-S-14876 \citep{Barro2016} (plus) and of GOODS-S-30274 \citep{Popping2017} (diamond), using the same color convention.}
    \label{fig: summary}
\end{figure*}

\begin{figure*}[ht]
    \centering
    \includegraphics[width = \textwidth]{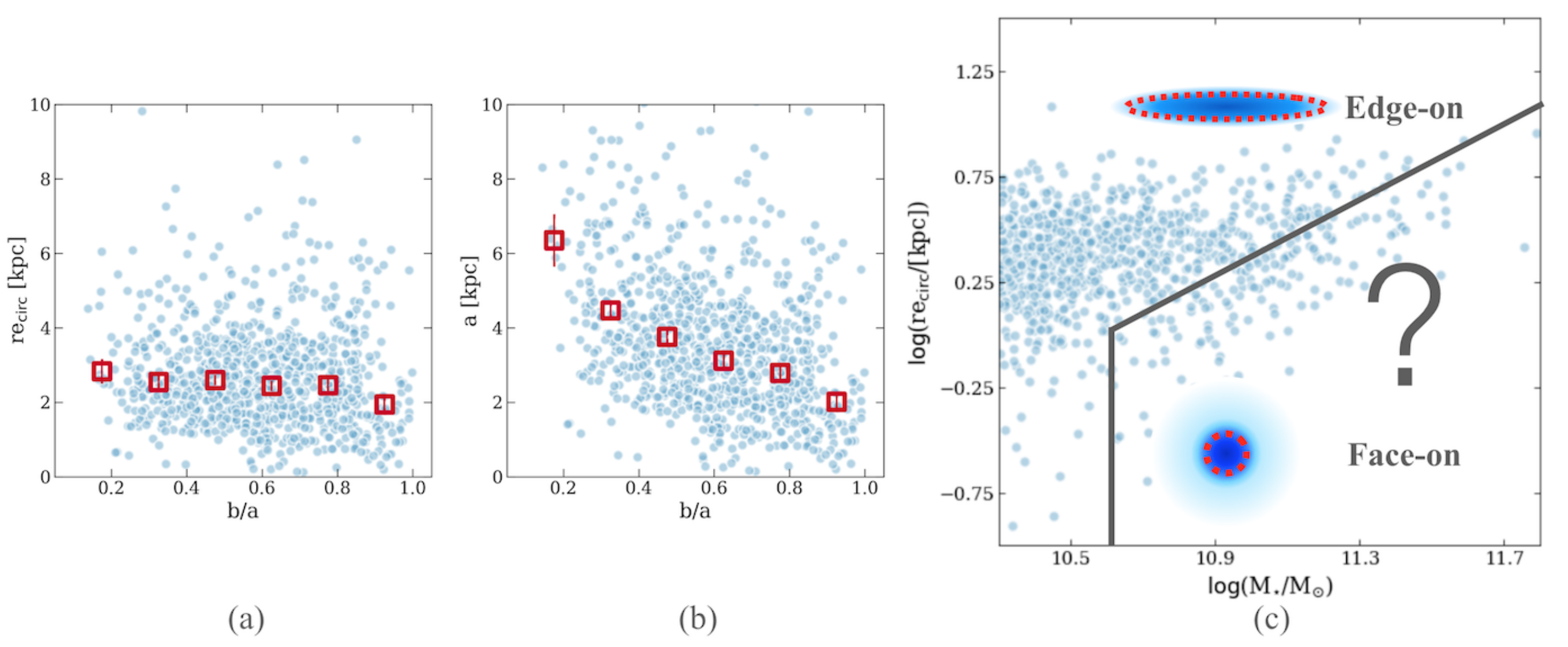}
    \caption{\textit{(a)} Circularized effective radius--axis ratio distribution, \textit{(b)} major axis effective radius--axis ratio distribution, and \textit{(c)} circularized effective radius--stellar mass distribution of star-forming galaxies with 10.5$<\log$(\mstar/\msun)$<$11.0 at 2.0$<z<$2.5 from \citet{VanderWel2014}. There is a strong correlation between $a$ and $b/a$. In \textit{(c)}, objects below the black lines were selected as compact massive star-forming galaxies in \citet{VanDokkum2015}. The red-dotted lines on cartoon galaxies represent their effective radius. The compact massive region of the size-mass plane may be biased towards face-on galaxies.}
    \label{fig:axisrat}
\end{figure*}



Even though we have large beam sizes ($\sim$ 1\farcs2) compared to the galaxy sizes ($r_e = $ 0\farcs1--0\farcs27), we investigated the spatially-resolved kinematics of the galaxies. Figure \ref{fig:kinematics} shows the observed rotation velocities for the three galaxies obtained by fitting the CO emission line at every spaxel with a single Gaussian. The velocity field of C2, which is our largest galaxy ($r_e=2.3$ kpc), reveals a continuous shear consistent with the kinematics of a rotating disk. Nothing conclusive can be said about either C1 or G5 and extent of the gas.
We derived the spatial extent of the molecular gas in C2 using CASA/\texttt{uvmodelfit}. We do a visibility fit of the emission in an averaged cube of width 200\,km\,s$^{-1}$ and find a half-light radius of 2.5$\pm$0.5 kpc by fitting a circular Gaussian disk. This is similar in size to the half-light radius of the galaxy (2.3 kpc; however this is measured from fitting a S\`ersic profile with n$=$3.3). C1 and G5 are not spatially resolved in our data. However, G5 was observed by \citet{Barro2016} at a much higher resolution (FWHM$\sim$ 0\farcs12-0\farcs18) using ALMA 870 $\mu$m dust continuum imaging. They find an effective radius of 1.14 kpc and a S\`ersic index of 0.7. A more detailed comparison between the spatial extent of stellar and dust disks will be performed in a future paper.

\section{Discussion} \label{sec:discussion}

In this {\em Letter} we tested the hypothesis that smaller-than-expected H$\alpha$ line widths of compact massive star forming galaxies at $z=2.0-2.5$ are caused by dust obscuration of their central regions. If this were the case, we would have measured broad, potentially double-horned, ALMA CO (3-2) emission lines. However, the measured line widths are FWHM$\sim$165$\pm$15 \kms\ in CO (3-2) for all three galaxies. These linewidths are much smaller than the 450 - 700 \kms\ we were expecting  and even
smaller than the H$\alpha$ line widths. We conclude that we can reject the hypothesis that the narrow H$\alpha$ lines in these objects were the result of preferential attenuation of the gas with the highest velocities.

We consider several possible alternative explanations. 
First, our observations are in the CO(3-2) line, and it has been shown that different CO transitions (reflecting different excitation levels) can trace gas with different spatial and kinematic properties \citep[e.g.,][]{Hodge2012,Bothwell2013,Casey2018}. There is some evidence that this may be relevant in COSMOS-27289, as \citet{Spilker2016} find that the CO(1-0) line has a FWHM of only 60\,\kms\ in this galaxy, almost a factor of three lower than our CO(3-2) measurement. However, other galaxies show similar kinematics for different tracers \citep[e.g.,][]{Popping2017}, and CO(3-2) should probe the denser and more turbulent gas. Nevertheless, it would be worthwhile to obtain higher transitions \citep[such as CO(8-7); see][]{Barro2017b}.

Second, the central regions of these galaxies could be devoid of (molecular and ionized) gas. In that case the ``missing'' high velocity H$\alpha$-emitting gas would not be hiding behind dust but simply does not exist, as the dense centers have already quenched. This interpretation finds some support in the fact that we spatially-resolve the CO gas in at least one of the galaxies (COSMOS-27289) \citep[see][]{Spilker2019}. However, it is difficult to reconcile with the rest-frame UV--optical morphology and spectral energy distributions of the galaxies. Their measured rest-frame optical sizes are small, and their rest-frame colors place them firmly in the ``dusty star forming" region of the $UVJ$ diagram (see vD15). It is difficult to imagine a scenario where the $150-400$\,\msun\,yr$^{-1}$ star formation in these galaxies only takes place outside of the visible regions {\em and} the central regions are misclassified in the $UVJ$ diagram and actually quiescent.

The third and perhaps most likely explanation is that the gas in these three galaxies is in rotation-dominated disks that we are viewing close to face-on. Our measured line widths are in good agreement with the intrinsic velocity dispersions estimated by \citet{Tadaki2017b} and \citet{Barro2017b} for similar galaxies and with the disk dispersions derived by \citet{Wisnioski2015} and \citet{Ubler2017} for the general population of massive star forming galaxies at $z\sim2-2.5$ \citep[see also][]{Kassin2014}. The narrow CO(1-0) line width of COSMOS-27289 measured by \citet{Spilker2016} is also consistent with this picture. Adopting the same excitation ratio between CO (3-2) and CO (1-0) and the same CO-to-H$_2$ conversion factor, the molecular gas fractions and depletion times are also consistent with those of the general population of star forming galaxies in the PHIBBS survey \citep{Tacconi2018} and the ASPECS program \citep{Aravena2019}. The issue with this explanation is that even though the axis ratios of the galaxies are high at $b/a\approx 0.8$, they are close to the median of the vD15 sample \citep[see also][]{Wisnioski2018}. It is therefore unlikely that galaxies with $b/a\approx 0.8$ are nearly face-on, {\em unless the compact massive region of the size-mass plane is bid towards face-on galaxies}.

We explore this possibility in Fig. \ref{fig:axisrat}a, which shows the relation between circularized size and axis ratio for massive  star forming galaxies. There is indeed a correlation, with a significance of 4$\sigma$, such that the smallest galaxies are on average rounder than the largest galaxies. The correlation is weak, but this is a reflection of the fact that we use the circularized size. If the galaxies are disks under random viewing angles, the relevant test is whether the major axis size, $a$, correlates with the axis ratio. This relation is shown in Fig.\ref{fig:axisrat}b. For an unbiased population of thin disks of different sizes there should be no correlation between these parameters: small and large disks would each show a uniform distribution of $b/a$. There is, however, a very strong correlation, with galaxies with $b/a\approx 0.1$ having a major axis that is a factor of $\sim 4$ larger than galaxies with $b/a\approx 0.9$. Although this relation could partially or entirely reflect differences in the morphology of large and small galaxies \citep[see, e.g.,][]{Zhang2019}, it is striking that there are no face-on counterparts to the large edge-on galaxies.

Such biases could have a variety of causes. Face-on disks have a lower stellar surface density than edge-on disks, whereas the stellar surface density of central bulge-like structures is less dependent on viewing angles. Combined with the generally low S/N ratio of the data and the use of single S\`ersic fits with a single axis ratio, this may lead to systematic underestimates of the sizes of face-on galaxies and overestimates of edge-on galaxies. Dust may also play a role. If there is significant dust in the disk, the central regions of edge-on galaxies will be more obscured than in face-on galaxies, and if the center is very dense this means face-on galaxies will appear to be more compact than edge-on galaxies \citep[see also][]{Graham2008,Gadotti2010,Price2017}.
Regardless of the cause, if there is an inclination bias in the size-mass plane it means that a significant fraction of apparently-compact star forming galaxies may in fact be face-on counterparts of galaxies with larger apparent sizes. This may be a common feature of samples that are selected to have small sizes, which would have important implications for the interpretation of the size-mass plane \citep{Mowla2019}. 


We note that these three galaxies have differences among themselves. The most massive and bluest galaxy COSMOS-27289 is almost three times as large and  has a lower inferred central density than the other two galaxies. While both COSMOS-27289 and COSMOS-10104 exhibit light in the outskirts, which can possibly be due to a disk, GOODS-S-5981 has extremely compact light profile. These suggest that these galaxies likely have different formation histories and may take different paths going forward.

To conclude, this study highlights the complexity of massive early star forming galaxies, and the need of dynamical information to understand their properties. It will be interesting to look for compact face-on quiescent disks at $z\sim1-2$, as our results suggest these maybe fairly common. Further progress can also be made with high resolution dust continuum imaging: if the dust continuum sizes do not show a dependence on axis ratio it would be strong evidence that the rest-frame optical sizes are biased. Early results have already suggested that galaxies tend to be smaller in dust continuum than in the rest-frame optical \citep[e.g.,][]{Tadaki2017a,Nelson2019}, in apparent conflict with the idea that the centers may be devoid of gas (the second explanation for the small line widths offered above). Ultimately deep rest-frame $3-4\,\mu$m imaging with JWST will show the actual morphologies of the galaxies, largely free from the effects of young stars and dust. Based on the results presented here these mid-IR sizes and morphologies may be quite different from the rest-frame optical ones that we have worked with so far.

\acknowledgements

We thank Arjen van der Wel and Emily Wisnioski for valuable feedback. We also thank the anonymous referee for insightful comments which improved our manuscript. LM thanks Héctor Arce, William Cramer and the CASA Helpdesk for helping to learn CASA. Support for this work was provided by NASA through the NASA Hubble Fellowship grant HST-HF2-51416.001-A awarded by the Space Telescope Science Institute, which is operated by the Association of Universities for Research in Astronomy, Inc., for NASA, under contract NAS5-26555.

\end{document}